\documentclass[10pt,letterpaper]{article}
\usepackage[top=0.85in,left=2.75in,footskip=0.75in]{geometry}

\usepackage{amsmath,amssymb}
\usepackage{changepage}
\usepackage[utf8x]{inputenc}
\usepackage{textcomp,marvosym}
\usepackage{cite}
\usepackage[right]{lineno}
\usepackage{microtype}
\DisableLigatures[f]{encoding = *, family = * }
\usepackage[table]{xcolor}
\usepackage{array}
\newcolumntype{+}{!{\vrule width 2pt}}
\newlength\savedwidth

\raggedright
\usepackage[aboveskip=1pt,labelfont=bf,labelsep=period,justification=raggedright,singlelinecheck=off]{caption}

\makeatletter
\renewcommand{\@biblabel}[1]{\quad#1.}
\makeatother

\date{}

\usepackage{lastpage,fancyhdr,graphicx}
\usepackage{epstopdf}
\fancyheadoffset[L]{2.25in}
\fancyfootoffset[L]{2.25in}
\begin{document}
\vspace*{0.2in}

\begin{flushleft}
{\Large
\textbf\newline{Punishment and inspection for governing the commons in a feedback-evolving game} 
}
\newline
\\
Xiaojie Chen\textsuperscript{1*} and
Attila Szolnoki\textsuperscript{2$\dagger$}
\\
\bigskip
\textbf{1} School of Mathematical Sciences, University of Electronic
Science and Technology of China, Chengdu 611731, China
\\
\textbf{2} Institute of Technical Physics and Materials Science, Centre for Energy Research, Hungarian Academy of Sciences, P.O. Box 49, H-1525 Budapest, Hungary
\\
\bigskip

* xiaojiechen@uestc.edu.cn\\
$\dagger$ szolnoki@mfa.kfki.hu

\end{flushleft}
\section*{Abstract}
Utilizing common resources is always a dilemma for community members. While cooperator players restrain themselves and consider the proper state of resources, defectors demand more than their supposed share for a higher payoff. To avoid the tragedy of the common state, punishing the latter group seems to be an adequate reaction. This conclusion, however, is less straightforward when we acknowledge the fact that resources are finite and even a renewable resource has limited growing capacity. To clarify the possible consequences, we consider a coevolutionary model where beside the payoff-driven competition of cooperator and defector players the level of a renewable resource depends sensitively on the fraction of cooperators and the total consumption of all players. The applied feedback-evolving game reveals that beside a delicately adjusted punishment it is also fundamental that cooperators should pay special attention to the growing capacity of renewable resources. Otherwise, even the usage of tough punishment cannot save the community from an undesired end.

\section*{Author summary}
Our proposed model considers not only the fundamental dilemma of individual and collective benefits but also focuses on their impacts on the environmental state. In general, there is a strong interdependence between individual actions and the actual shape of environment that can be described by means of a co-evolutionary model. Such approach recognizes the fact that even if our common-pool resources are partly renewable, they have limited growth capacities hence a depleted environment is unable to recover and reach a sustainable level again. This scenario would have a dramatic consequence on our whole society, therefore we should avoid it by punishing those who are not exercising restrain. We provide analytical and numerical evidences which highlight that punishment alone may not necessarily be a powerful tool to maintain a healthy shape of environment for the benefit of future generations. Cooperator actors, who are believed to take care of present state of our environment, should also consider carefully the growth capacity of renewable resources.


\section*{Introduction}

Overexploitation of common-pool resources is a fundamental problem that can be identified in several seemingly different ecological systems \cite{ostrom_90,hardin_g_s68}. A well-known example is the danger of overfishing. Fishermen are motivated to catch the maximum amount of fish because restraint could only work if all others are behaving similarly. Otherwise, fish are driven to extinction which is the worst scenario for everyone \cite{pauly_n02,kraak_ff11}. Similarly, we can continue this list endlessly by giving further examples, like overgrazing of common pasture lands, where individual short-term benefit seems to be in conflict with long-term interest of a larger population. The mutual feature of these cases is human activity influences the actual state of resources which has a negative feedback for not only those who degrade the environment but also for the whole community. We stress that this problem is not restricted to human-related activities, but may also appear at microscopic level including microbes, bacterias, and viruses \cite{crespi_tee01, hummert_jtb10, west_prslb03,rankin_tee07,gore_n09, schuster_bs11}, which explains why the problem of common-pool resource exploitation is an intensively studied research area of several disciplines \cite{cox_es10,hauser_n14,sugiarto_prl17}.

It is a fundamental point that the sustainable use of common-pool resources is strongly based on the interdependence of resource and social dynamics \cite{brander_aer98}. On the one hand, the dynamics of resources, in particular renewable resources, are influenced by some ecological factors, such as the resource growth rate and the carrying capacity \cite{brander_aer98,tavoni_jtb12}. On the other hand, these resources are also influenced by human behaviors on how to use them. Meanwhile, the shape of a dynamical resource also influences the prosperity of human well-being, which triggers frequency-dependent changes in individual strategies \cite{weitz_pnas16}. Thus, the interaction of resource dynamics and the evolution of individual-based behavior can be captured properly by a feedback-evolving game model where both variables are in the focus of governing equations \cite{tavoni_jtb12,lade_te13,schluter_prsb16,weitz_pnas16,lee_jtb17}.

A frequently recommend solution for sustaining the requested level of common-pool resource could be to punish defectors for over-harvesting \cite{fehr_aer00,henrich_s06,gachter_s08,sigmund_n10,vasconcdlos_ncc13,pacheco_plr14,johnson_rsos15,chen_jrsi15,chen_srep15,okada_ploscb15,perc_pr17,wang_amc18}. In parallel, some other related control mechanisms, like ostracism or voluntary enforcement, are also discussed as viable solutions to the original problem \cite{nakamaru_pone14,sugiarto_jpc17}. Importantly, the consequence of top-down regulation, which is based on inspection, permanent monitoring of agents and punishment, has been used for the forest commons management \cite{yang_pnas13}, but still begs for clarification especially in the presence of renewable resources.

Thus, in this work our principal interest is to explore how the application of punishment and inspection influences the competition of strategies when the benefit of given strategies depends sensitively on the actual state of environment. Furthermore, in our approach the common resource is considered as a dynamically renewable system which is also influenced by a feedback of individual strategies. This interdependence can be modeled by a co-evolutionary system where both strategies and environmental resources are subject to change. In our work, we depict this interdependent relationship by using the replicator equation for the evolution of strategies and the logistic growth model for the resource dynamics \cite{hofbauer_98,tsoularis_mb02}, which provides a novel approach in the field of socio-ecological dynamics, to our knowledge. Indeed, some theoretical models have already raised the concept of environmental coupling \cite{tavoni_jtb12, brandt_pone12,chen_srep14,weitz_pnas16}, but our present approach considers the growing capacity of a renewable resource explicitly. Additionally, punishment and permanent monitoring (inspection) are used as a control mechanism for blocking overexploitation of the common resource. We demonstrate that in addition to a delicately adjusted punishment regime, the growing capacity of renewable resource is fundamental for keeping resources at a sustainable level.

\section*{Materials and methods}

We consider a population of individuals who all use a common resource at different levels. While the resource amount $y$ in the
common pool is limited, we assume that it is partly renewable and its dynamics can be described by the frequently used logistic
population growth model \cite{hofbauer_98,tsoularis_mb02}. Accordingly, the dynamical change of the resource amount induced by
its environment factor is given by $\dot{y}=ry(1-y/R_m)$, where $r$ is the intrinsic growth rate and $R_m$ is the carrying capacity of
resource pool. Furthermore, for the proper utilization of the resource, individuals are allocated an amount of resource from the
common pool, which depends on the total resource amount $y$ at a given time. According to the allocation rule, we suppose that the
legal amount that each individual can be allocated from the pool is $b_l = b_m y /R_m$, where $b_m$ is the maximal resource portion that
each individual is allowed to use per unit of time when the amount of the common pool $y$ reaches $R_m$. Evidently, this individual limit portion satisfies $b_m \leq R_m / N$, where $N$ is the total number of individuals in the community.

For simplicity, we assume that individuals choose between two basically different strategies. The first group is called as
cooperators or ``law-abiding" individuals who follow the allocation rule and restrain their use to $b_l$ amount from the resource. The
other group, called as defectors or ``violators", ignores the allocation rule and utilizes the common pool more intensively by
getting a larger $b_v>b_l$ amount. Here we suppose that $b_v=b_l(1+\alpha)$, where $\alpha>0$ characterizes the severity of
defection.

In order to avoid resource exploitation, we introduce a centrally organized inspection and punishing mechanism often used in realistic resource management systems \cite{yang_pnas13,chen_srep15}. In particular, we assume that defection is detected with a probability $p$ ($0<p<1$), which is the probability of detecting a defector during a time unit. If an individual is identified for overexploiting common resources, then it will be punished with a fine $\beta$ ($\beta>0$) which is reduced from his collected payoff. In this way, the parameter $p$ characterizes the effectiveness of monitoring system while the parameter $\beta$ describes how severe the applied punishment.

To explore the possible impact of inspection and punishment on the evolutionary process, we consider a well-mixed population and employ
the replicator equation that describes the time evolution of competing strategies \cite{hofbauer_98}. If we denote the fraction
of cooperators by $x$ then the governing equation is
\begin{eqnarray}
\dot{x}=x(1-x)(P_L-P_V),\label{eq1}
\end{eqnarray}
where $P_L$ and $P_V$ are the payoffs of law-abiding individual (cooperator) and violator (defector) players, respectively. Notice that a player's income is originated from the common-pool resource, although there is a coupling between individual payoff obtained from the feedback-evolving game and resource dynamics \cite{tavoni_jtb12}. Thus, in our study for simplicity we assume that the related payoff values are directly written as $P_L=b_l$ and $P_V=b_v-p\beta$ for cooperators and defectors, respectively.

In agreement with the co-evolutionary concept which considers the feedback of individual acts on the actual state of resource, the
governing equation of common resource abundance can be extended as follows
\begin{eqnarray}
\dot{y}=ry(1-y/R_m)-N[b_l x + (1-x) b_v]. \label{eq2}
\end{eqnarray}

In the next section we analyze and discuss the possible equilibrium points of the above coupled equation system. Furthermore we extend
our study by presenting the results of individual-based Monte Carlo simulations as a supplement to support the validity of our mathematical analysis for wider conditions.

\section*{Results}

\subsection*{Equilibrium states of the feedback-evolving dynamical system}

By substituting the payoff values into Eq.~\ref{eq1}, we have the following equation system
$$ \left\{
\begin{aligned}
\dot{x} & = x(1-x)(p\beta-\frac{y}{R_m}b_m\alpha) \\
\dot{y} & =ry(1-y/R_m)-N\frac{y}{R_m}b_m[1+(1-x)\alpha].
\end{aligned}
\right.
$$

This equation system has at most five fixed points which are $[x, y]=[0, 0]$, $[0, R_m-\frac{b_m N(1+\alpha)}{r}]$,  $[1,
0]$, $[1, R_m-\frac{Nb_m}{r}]$, and
$[1+\frac{1}{\alpha}-\frac{R_mr}{\alpha b_m N}+\frac{p\beta
R_mr}{{\alpha}^2 b^2_m N}, \frac{p \beta R_m}{b_m\alpha}]$, respectively. Here the first four are boundary fixed points, while the last one is an interior fixed point. By calculating the first order partial derivaties \cite{khalil_12}, the Jacobian matrix of our system can be written as

\begin{eqnarray*}
\begin{array}{cc}
J=\left[\begin{array}{cc} (1-2x)(p\beta-\frac{\alpha b_m y}{R_m}) & \frac{\alpha b_m x(x-1)}{R_m} \\
\frac{\alpha b_m N y}{R_m} & r-\frac{b_mN(1+\alpha-\alpha
x)+2ry}{R_m}\end{array}\right].
\end{array}
\end{eqnarray*}

The specific form of this matrix at the above mentioned fixed points are
\begin{eqnarray*}
\begin{array}{cc}
J(0,0)=\left[\begin{array}{cc} p\beta & 0 \\
0 & r-\frac{b_mN(1+\alpha)}{R_m}\end{array}\right],
\end{array}
\end{eqnarray*}

\begin{eqnarray*}
\begin{array}{cc}
J(1,0)=\left[\begin{array}{cc} -p\beta & 0 \\
0 & r-\frac{b_mN}{R_m}\end{array}\right],
\end{array}
\end{eqnarray*}

\begin{eqnarray*}
\begin{array}{cc}
J(0,R_m-\frac{Nb_m(1+a)}{r})=\left[\begin{array}{cc} p\beta+\frac{\alpha b_m[(1+\alpha)b_mN-R_mr]}{R_mr} & 0 \\
\frac{\alpha b_mN[R_mr-(1+\alpha)b_mN]}{R_m r} &
\frac{(1+\alpha)b_mN-R_mr}{R_m}\end{array}\right],
\end{array}
\end{eqnarray*}

\begin{eqnarray*}
\begin{array}{cc}
J(1,R_m-\frac{Nb_m}{r})=\left[\begin{array}{cc} \frac{\alpha b_m(R_mr-b_mN)}{R_mr}-p\beta & 0 \\
\frac{\alpha b_mN(R_mr-b_mN)}{R_m r} &
\frac{b_mN-R_mr}{R_m}\end{array}\right],
\end{array}
\end{eqnarray*}

and

\begin{eqnarray*}
\begin{array}{cc}
J(1+\frac{1}{\alpha}-\frac{R_mr}{\alpha b_m N}+\frac{p\beta
R_mr}{{\alpha}^2 b^2_m N}, \frac{p \beta
R_m}{b_m\alpha})=\left[\begin{array}{cc} 0 &
(\frac{1}{\alpha}-\frac{R_mr}{\alpha b_m N}+\frac{p\beta
R_mr}{{\alpha}^2 b^2_m N})[\frac{b_m(1+\alpha)}{R_m}-\frac{r}{N}+\frac{p\beta r}{\alpha b_m N}] \\
Np\beta & -\frac{p\beta r}{\alpha b_m}\end{array}\right].
\end{array}
\end{eqnarray*}

The stability of these fixed points can be determined from the sign of the eigenvalues of the Jacobian \cite{ khalil_12}. It is easy to see that the eigenvalues of the matrices for the boundary fixed points are the corresponding diagonal elements. Hence, the stability of these fixed points depend exclusively on the signs of the diagonal elements of the related matrices. In addition, the trace of the last Jacobian is negative, hence the interior fix point has at least one negative eigenvalue. It also involves that the stability of the unique interior fixed point depends only on the sign of $(\frac{1}{\alpha}-\frac{R_mr}{\alpha b_m N}+\frac{p\beta R_mr}{{\alpha}^2 b^2_m N})[\frac{b_m(1+\alpha)}{R_m}-\frac{r}{N}+\frac{p\beta r}{\alpha b_m N}]$ term. For further analysis let us denote $e_c=\frac{b_mN}{R_m}$ and $e_d=\frac{b_mN(1+\alpha)}{R_m}$ respectively, representing the gain rates of cooperators and defectors in a population from the common resource. It also involves that we have $0<e_c\leq 1$ and $e_c<e_d$.

In the following, we distinguish three substantially different parameter regions where the distinction is based on the actual intrinsic growth rate value of the renewable common pool resource.

\subsubsection*{Slowly growing resource pool}

First we consider the case when the resource pool is recovering slowly due to small intrinsic grow rate, which assumes that $0<r<e_c<e_d$. In this situation, we have $r R_m < b_m N$, and accordingly the system has only two fixed points in the parameter space of $0\leq x\leq 1$ and $y\geq 0$. They are $[0, 0]$ and $[1, 0]$, respectively. Here the largest eigenvalue of $J(0,0)$ is positive, whereas the largest eigenvalue of $J(1,0)$ is negative due to the small value of $r$. Consequently, the fixed point $[0,0]$ is unstable, while the fixed point $[1,0]$ is stable. For the special case of $r=e_c$, we find that one eigenvalue of the Jacobian matrix at the fixed point $[1,0]$ is zero and the other one is negative. In the SI text, we further provide the stability analysis of the fixed point by using the center manifold theorem \cite{khalil_12}.

\begin{figure}[!h]
\begin{center}
\includegraphics[width=5in]{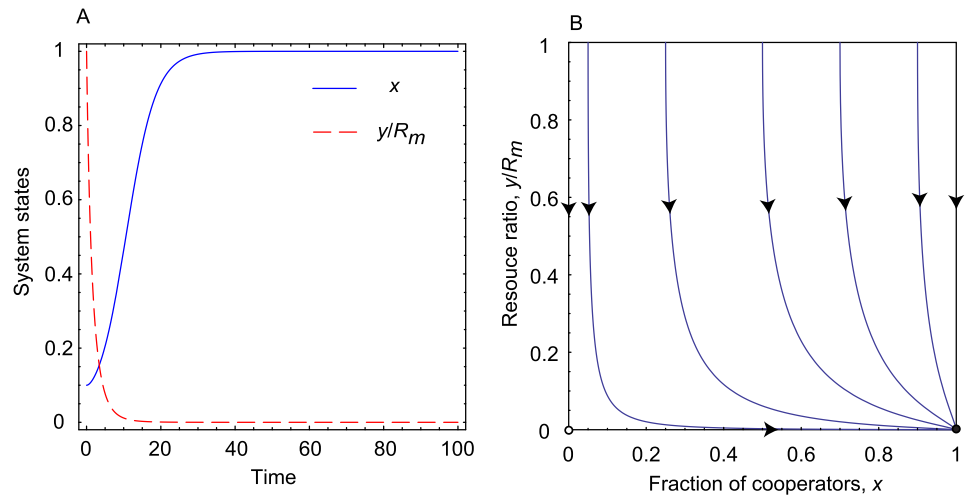}
\end{center}
\caption{{\bf Replicator dynamics for slowly growing resource.}
Panel A: Time evolution of the fraction of cooperators and the resource abundance ratio. Panel B: Phase portrait on $x-y/R_m$ plane for $r<e_c<e_d$. Filled circle represents a stable fixed point, while open circle represents an unstable fixed point. For $r<e_c<e_d$ the cooperation level is satisfactory in the final state, but resource abundance becomes depleted due to low growing resource rate. Parameters are $r=0.3$, $N=1000$, $R_m=1000$, $p=0.5$, $\alpha=0.5$, $\beta=0.5$, and $b_m=0.5$.}
\label{fig1}
\end{figure}

A representative time evolution of the cooperation level and the abundance of common resource pool for $0<r<e_c<e_d$ is plotted in Fig.~\ref{fig1}. It suggests that while the inspection and punishing mechanisms are capable to drive the system toward a full cooperator state, but this destination remains still unsatisfactory because the resource pool becomes fully depleted. This result warrants that it is not enough to be cooperator and consider only the actual shape of common resource pool. If the intrinsic growing rate of the latter is too low, then users should take a much lower share from the pool than it is believed naively based on the present status. Otherwise the common resource pool is unable to renew and the high cooperation level becomes useless.

\subsubsection*{Moderately growing resource pool}

\begin{figure}[!h]
\begin{center}
\includegraphics[width=5in]{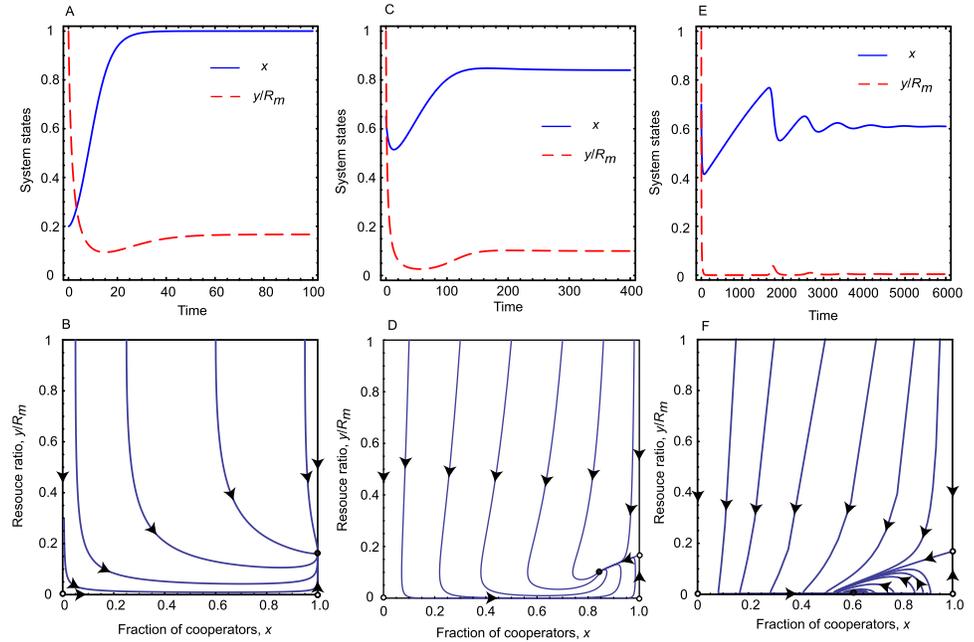}
\end{center}
\caption{{\bf Replicator dynamics for moderately growing resource.} Top panels show the time evolution of the fraction of cooperators and the resource abundance ratio for different parameter values when $e_c<r<e_d$. Bottom panels show the related phase portraits on $x-y/R_m$ plane. Parameters for Panels A and B: $r=0.6$, $N=1000, R_m=1000$, $p=0.5$, $\alpha=0.5$, $\beta=0.5$, and $b_m=0.5$; for Panels C and D: $r=0.6$, $N=1000$, $R_m=1000$, $p=0.05$, $\alpha=0.5$, $\beta=0.5$, and $b_m=0.5$; for Panels E and F: $r=0.6$, $N=1000$, $R_m=1000$, $p=0.01$, $\alpha=0.5$, $\beta=0.1$, and $b_m=0.5$. These plots suggest that a sustainable state can be reached for appropriate inspection and punishment level, but the depleted environment state cannot be avoided if these institutions are ineffective.}
\label{fig2}
\end{figure}

If the intrinsic growth rate of resource pool is moderate, means $0 < e_c < r < e_d$, then the conclusion is more subtle. In this situation, we have $b_m N < r R_m < b_m N (1+\alpha)$ and $1+\frac{1}{\alpha}-\frac{R_m r}{\alpha b_m N}+\frac{p \beta R_m r}{{\alpha}^2 b^2_m N}>0$. In dependence of the efficiency of inspection and punishment we can distinguish two main cases. Note that the combined effect of these institutions can be characterized by the product of $p$ and $\beta$ parameters. The first case is when they are efficient hence their product exceeds $p\beta>b_m\alpha(1-\frac{e_c}{ r})$. In this case $1+\frac{1}{\alpha}-\frac{R_mr}{\alpha b_m N}+\frac{p\beta
R_mr}{{\alpha}^2 b^2_m N}>1$ is also fulfilled. As a result, the system has three fixed points which are $[0, 0]$, $[1, 0]$, and $[1, R_m -\frac{N b_m}{r}]$, respectively. According to the sign of the largest eigenvalues $[0,0]$ and $[1, 0]$ are unstable, while $[1,R_m-\frac{Nb_m}{r}]$ is a stable fixed point. This result suggests that the system reaches an equilibrium point where all participants share the common pool cooperatively and the renewable resource is capable to maintain a sustainable level. This case is illustrated in the first column of Fig.~\ref{fig2}.

When the inspection-punishment institutions are less effective, then the term $b_m \alpha (1 - \frac{e_c}{r})$ exceeds $p \beta$ products. In this case $1 + \frac{1}{\alpha} - \frac{R_m r}{\alpha b_m N}+\frac{p \beta R_m r}{{\alpha}^2 b^2_m N}< 1$ and the system has four fixed
points. They are $[0, 0]$, $[1, 0]$, $[1, R_m - \frac{N b_m}{r}]$, and $[1+\frac{1}{\alpha}-\frac{R_mr}{\alpha b_m N}+\frac{p\beta R_m r}{{\alpha}^2 b^2_m N}, \frac{p \beta R_m}{b_m\alpha}]$, respectively. Here only the last fixed point is stable while the first three are unstable. In equilibrium cooperators and defectors coexist at a finite resource abundance which is inversely proportional to $\alpha$ that characterizes how intensively defector players over exploit the common resource. The equilibrium resource level is linearly proportional to the $p \beta$ product, while the first part of the equilibrium fixed point contains a term which is free from $p \beta$. This means that the stable fixed point drifts toward $y=0$ axis faster than to the $x=0$ axis as we weaken the impact of inspection-punishment institutions. Consequently, when the impact of inspection and punishment tends to zero-limit then the system still remains in a mixed state of cooperators and defectors but it has no particular importance because the environment becomes depleted. This is illustrated in the right column of Fig.~\ref{fig2} where the stable fixed point approached the horizontal axis as we lower $p$ that is the probability of successful detection of overexploitation.

In the special case when $p\beta=b_m\alpha(1-\frac{e_c}{ r})$, we find that there are three fixed points in the system, which are $[0,0]$, $[1,0]$, and $[1, R_m-\frac{Nb_m}{r}]$, respectively. But one eigenvalue of the Jacobian matrix at the fixed point $[1, R_m-\frac{Nb_m}{r}]$ is zero and the other one is negative. In the SI text, we provide the stability analysis of the fixed point by using the center manifold theorem \cite{khalil_12}. Furthermore, we provide the theoretical analysis of the equilibrium points for the special case of $r=e_d$.

\subsubsection*{Rapidly growing resource pool}

\begin{figure}[!h]
\begin{center}
\includegraphics[width=5in]{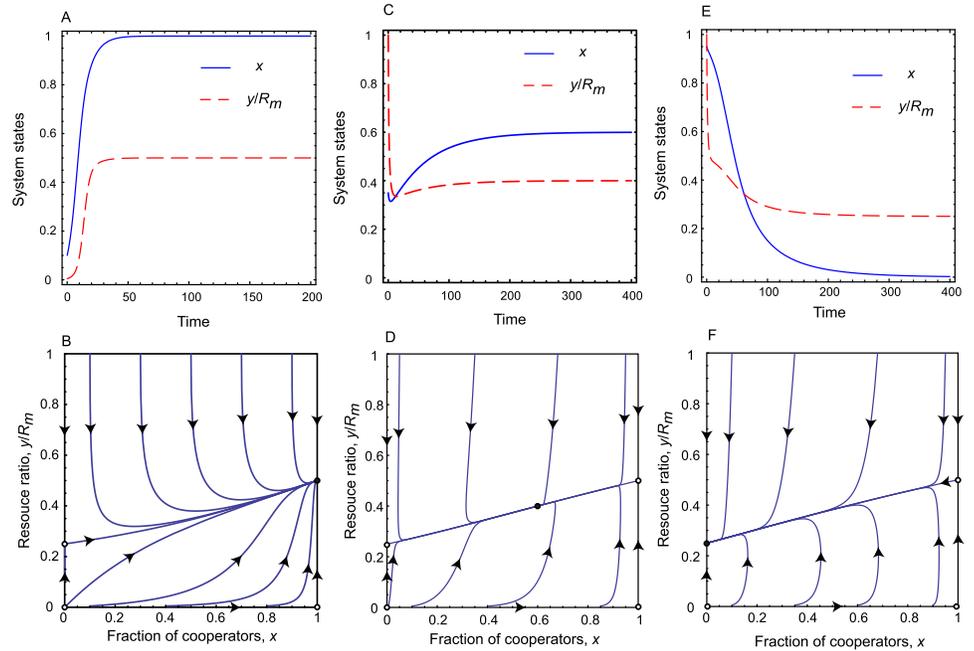}
\end{center}
\caption{{\bf Replicator dynamics for rapidly growing resource.} Top panels show the time evolution of the fraction of cooperators and the resource abundance ratio for different parameter values when $e_c<e_d<r$. Bottom panels show the related phase portraits on $x-y/R_m$ plane. Parameters for
 Panels A and B: $r=1.0$, $N=1000$, $R_m=1000$, $p=0.5$, $\alpha=0.5$, $\beta=0.5$, and $b_m=0.5$; for Panels C and D: $r=1.0$, $N=1000$, $R_m=1000$, $p=0.2$, $\alpha=0.5$, $\beta=0.5$, and $b_m=0.5$; for Panels E and F: $r=1.0$, $N=100$, $R_m=100$, $p=0.1$, $\alpha=0.5$, $\beta=0.5$, and $b_m=0.5$. Due to the large intrinsic growth rate, the environmental resource will never be depleted. The strength of external institutions determines the relation of competing strategies in the equilibrium state.}
\label{fig3}
\end{figure}

Finally we discuss the case when the intrinsic growth rate of resource is large enough to exceed both the gain rate of cooperators $e_c$ and the gain rate of defectors $e_d$. Here, we have $rR_m>b_mN$ and $rR_m>b_mN(1+\alpha)$. As previously, we can distinct two significantly different cases depending on the power of inspection and punishment institutions. If they are strong enough and the product of $p \beta$ exceeds $b_m\alpha(1-\frac{e_c}{r})$, then we have $1+\frac{1}{\alpha}-\frac{R_mr}{\alpha b_m N}+\frac{p\beta
R_mr}{{\alpha}^2 b^2_m N} > 1$. In this case the system has four fixed points which are $[0, 0]$, $[1, 0]$, $[0,R_m-\frac{b_m N(1+\alpha)}{r}]$, and $[1, R_m-\frac{Nb_m}{r}]$, respectively. While the first three are unstable, the last one is a stable fixed point. This means that the system  evolves into a full cooperator state where environmental resource stabilizes at a sustainable level. This level depends practically on the growth rate of resource. A representative phase portrait is plotted in the left column of Fig.~\ref{fig3}.

If the above mentioned institutions are less powerful, then the product $p \beta$ cannot beat $b_m\alpha(1-\frac{e_c}{r})$ value. Hence we have $1+\frac{1}{\alpha}-\frac{R_mr}{\alpha b_m N}+\frac{p\beta R_mr}{{\alpha}^2 b^2_m N} < 1$. Depending on the actual strength of inspection and punishment we can distinguish two subcases here. First, when the above mentioned institutions are still considerable, $p \beta$ exceeds $b_m\alpha(1-\frac{e_d}{ r})$ and the term $1+\frac{1}{\alpha}-\frac{R_mr}{\alpha b_m N}+\frac{p\beta
R_mr}{{\alpha}^2 b^2_m N}$ is positive. As a result, the system has five fixed points which are $[0, 0]$, $[0, R_m-\frac{b_m N(1+\alpha)}{r}]$,  $[1, 0]$,
$[1, R_m-\frac{Nb_m}{r}]$, and
$[1+\frac{1}{\alpha}-\frac{R_mr}{\alpha b_m N}+\frac{p\beta
R_mr}{{\alpha}^2 b^2_m N}, \frac{p \beta R_m}{b_m\alpha}]$, respectively. Here only the last fixed point is stable while all the others are unstable. This scenario is illustrated in the middle column of Fig.~\ref{fig3}. From this result we can conclude that a reasonably strong external institution is capable to maintain the coexistence of cooperator and defectors states. Their fractions depend principally on the difference between resource usages of strategies which is characterized by the parameter $\alpha$.

According to the second subcase, when the institutions are too weak and $p \beta$ product cannot exceed $b_m\alpha(1-\frac{e_d}{r})$, the term $1+\frac{1}{\alpha}-\frac{R_mr}{\alpha b_m N}+\frac{p\beta
R_mr}{{\alpha}^2 b^2_m N}$ is negative. In this case the system has four fixed points which are $[0, R_m-\frac{b_m N(1+\alpha)}{r}]$,  $[0, 0]$, $[1, 0]$,
and $[1, R_m-\frac{Nb_m}{r}]$, respectively. Here only the first fixed point is stable, while the rest are unstable. In this situation, which is illustrated in the right column of Fig.~\ref{fig3}, the evolution terminates into a full defection state. Still the latter is a sustainable state because the strong growing capacity of resource is capable to compensate to greediness of defector players.

\begin{figure}[!h]
\begin{center}
\includegraphics[width=5in]{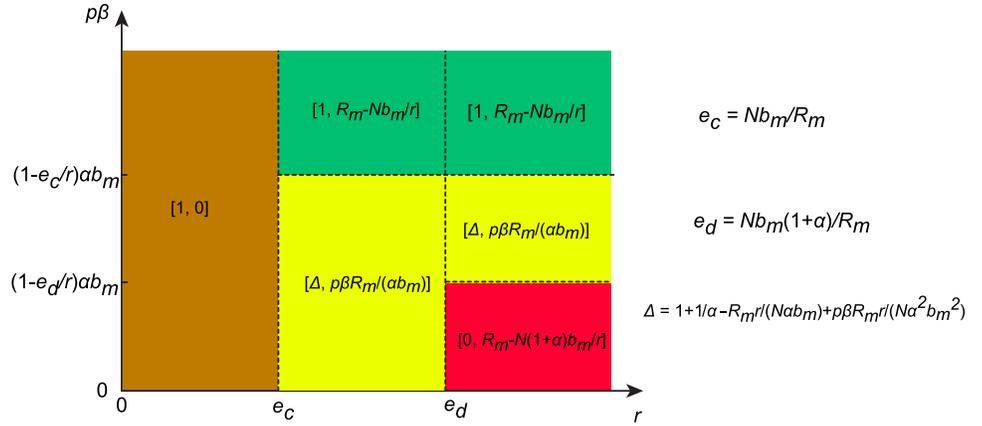}
\end{center}
\caption{{\bf A representative plot of evolutionary outcomes on the phase plane.} Different colors are used to distinguish qualitatively different solutions in the parameter space ($r$, $p\beta$). This plot highlights that the inner dynamical feature of renewable resource could be a
decisive factor that can derogate the expected consequence of punishment.}
\label{fig4}
\end{figure}

Finally, we point out that there exist two special cases of $\alpha b_m(1-\frac{e_c}{r})=p\beta$ and $\alpha b_m(1-\frac{e_d}{r})=p\beta$ for the rapidly growing resource pool situation. We provide the theoretical analysis for the equilibrium points in these two special cases in the SI text. In addition, in order to help readers to overview easily the evolutionary stable states of our feedback-evolving dynamical system for different parameter regions, we present an illustrative plot of the dynamical regimes in the parameter space ($r$, $p\beta$), as shown in Fig.~\ref{fig4}. We use different colors to distinguish the qualitatively different solutions for different parameter values.

\subsection*{Monte Carlo simulations}

To support the robustness of the predictions made by our presented mathematical analysis, we perform Monte Carlo simulations \cite{maciejewski_ploscb14,adami_plr16,wu_ploscb17} which may serve as an alternative approach to explore the possible coevolutionary dynamics. Indeed, in some cases this alternative approach, which contains stochastic elements and utilizes microscopic dynamics, goes beyond the limitations of macroscopic equations and provides alternative outcomes of evolutionary dynamics that are absent from a well-mixed behavior\cite{adami_plr16}. In the present case, however, this technique confirms our previous findings, and hence underlines the broader robustness of our observations.

In the Monte Carlo simulations, initially each individual in the population is chosen to violate or to follow the allocation rule. These strategies are denoted by $s_v=0$ and $s_l=1$ respectively. In agreement with the previous setup a law-abiding or cooperator player gets an amount $y(t)b_m/R_m$ from resource which provides his payoff value. Here $y(t)$ describes the actual state of resource pool at time step $t$. Alternatively, a defector who violates the allocation rule takes a larger amount $y(t)b_m/R_m(1+\alpha)$ from the common resource pool. However, the whole population is inspected and defection is identified with a probability $p$. In this case the identified defector player is punished and a fine $\beta$ is reduced from his payoff. Technically, it means that a defector collects a payoff $P_i=(1+\alpha)y(t)b_m/R_m - \beta$ with probability $p$, otherwise his final payoff is
$P_i=(1+\alpha)y(t)b_m/R_m$.

Because of the feedback mechanism the total amount of common resources is updated according to the rule
$$
y(t+1)=y(t)+ry(t)[1-\frac{y(t)}{R_m}]-\sum_i^{N}[s_i\frac{y(t)b_m}{R_m}+(1-s_i)\frac{y(t)b_m(1+\alpha)}{R_m}]\,,
$$
where both the intrinsic growth of resource and the exploitation effect are considered.

According to the strategy evolution each individual $i$ has a chance to imitate the strategy of another randomly chosen individual $j$. If $P_i<P_j$, then the strategy transfer occurs with the probability
$$
q=\frac{P_j-P_i}{M},
$$
where $M$ ensures the proper normalization and is given by the maximum possible difference between the payoffs of $i$ and $j$ players \cite{santos_n08}.

\begin{figure}[!h]
\begin{center}
\includegraphics[width=3in]{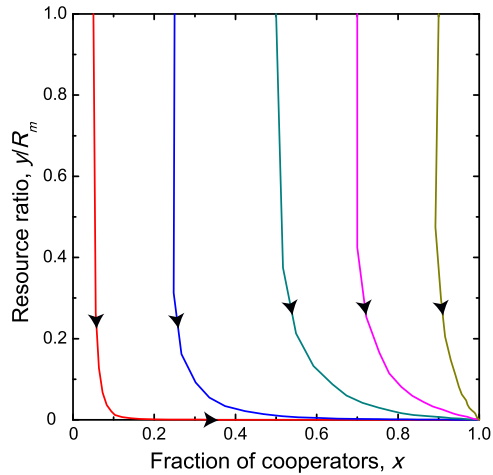}
\end{center}
\caption{{\bf Individual-based simulation for slowly growing resource.} Phase portraits in $x-y/R_M$ plane for different initial conditions when $r<e_c<e_d$. Simulation results show that the system converges to the $[1, 0]$ state regardless of the initial conditions. Parameters: $r=0.3$, $p=0.5$, $N=1000$, $R_m=1000$, $\alpha=0.5$, $\beta=0.5$, and $b_m=0.5$.}
\label{fig5}
\end{figure}

Our results obtained at slow resource growth are summarized in Fig.~\ref{fig5}. This representative plot confirms our previous observations, namely, the effective inspection and punishment institutions are not enough to reach a sustainable state if the resource grow rate is too small. In this case the mentioned institutions are capable to shift the system toward a full cooperator state but it does not help because even law-abiding users are myopic and only consider the actual state of environment. As a conclusion, exploiting resource based on the present shape is harmful because common resource cannot recover this seemingly suitable usage.

\begin{figure}[!h]
\begin{center}
\includegraphics[width=5in]{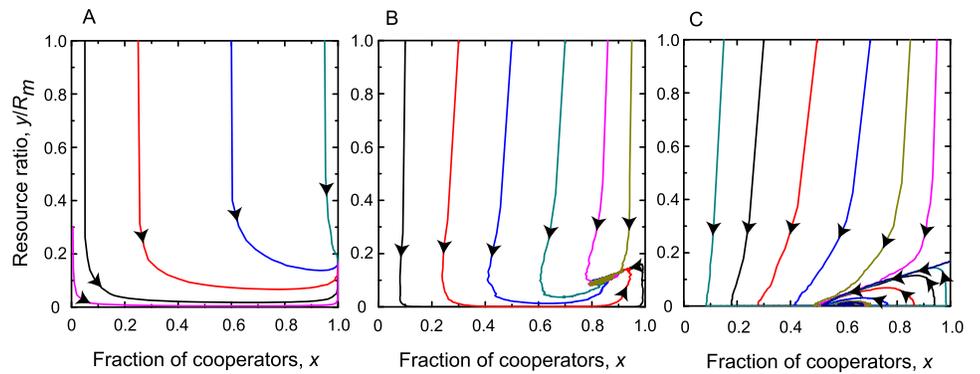}
\end{center}
\caption{{\bf Individual-based simulation for moderately growing resource.} Three representative phase portraits in $x-y/R_M$ plane using different initial conditions when $e_c<r<e_d$. Panels show results obtained at powerful (left), moderate (middle), and weak (right) external institutions. Depending on the effectiveness of inspection and punishment a sustainable state can be reached for the first two cases. The specific values are $p=0.5$ and $\beta=0.5$ in panel A; $p=0.05$ and $\beta=0.5$ in panel B; and $p=0.01$ and $\beta=0.1$ in panel C. Other parameters are $r=0.6$, $N=1000$, $R_m=1000$, $\alpha=0.5$, and $b_m=0.5$ for all cases.}
\label{fig6}
\end{figure}

When the intrinsic growth rate is higher, but moderate then we obtain qualitatively similar results to those obtained by analyzing the equation system. These observations are summarized in Fig.~\ref{fig6}. They suggest that in this case the application of inspection and punishment may result in the desired effect and a sustainable environmental state can be reached. But here the efficiency of applied top-down governance plays a decisive role in the final outcome.

\begin{figure}[!h]
\begin{center}
\includegraphics[width=5in]{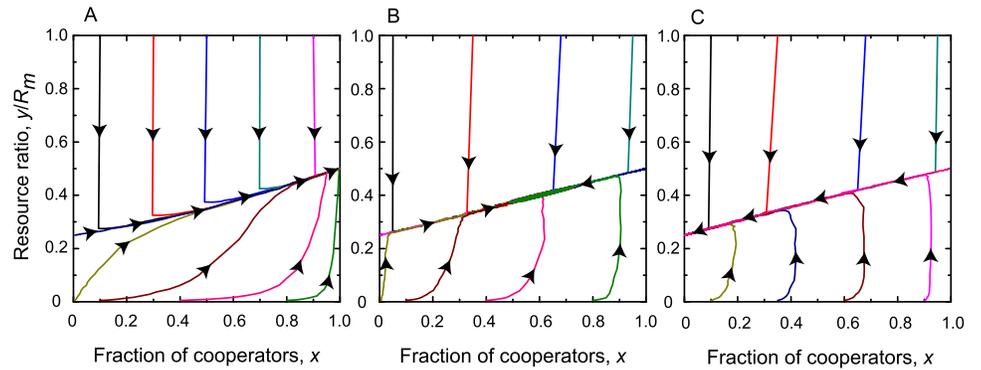}
\end{center}
\caption{{\bf Individual-based simulation for rapidly growing resource.}
Three representative phase portraits in $x-y/R_M$ plane using different initial conditions when $e_c<e_d<r$. Panels show results obtained at powerful (left), moderate (middle), and weak (right) external institutions. Due to large growth rate a sustainable state can always be maintained. The strength of inspection and punishment determines only the level of this state. The specific values are  $p=0.5, 0.2$, and $0.1$ for panel A, B, and C, respectively. Other parameters are $r=1.0$, $N=1000$, $R_m=1000$, $\alpha=0.5$, $\beta=0.5$, and $b_m=0.5$ for all cases. }
\label{fig7}
\end{figure}

Finally we summarize our observations in Fig.~\ref{fig7} obtained for the rapidly growing resource case. These results are again in agreement with the prediction of equation system. More precisely, full $D$ state, stable coexistence of $C$ and $D$ strategies, or full $C$ state can also be obtained in dependence of the strength of inspection and punishment. The latter, however, have only second order importance because the fast recovery of environmental resource pool is capable to maintain a sustainable state for all cases.

\section*{Discussion}

The surprising efficiency of evolutionary game theory in
understanding our complex world in widely different scales makes possible to predict the long-term consequences of individual actions on resource management \cite{tavoni_ncc13,szolnoki_plr14}. Our principal aim is to develop a realistic model where there is a coupling between the behavior of players and the developing state of a common pool resource. More precisely, we consider not just exploitation of the resource but also take account into the fact that the environmental common pool can be renewable. The latter may be captured by a single parameter which characterizes the intrinsic growth rate of the resource.

In this more realistic model of a coupled social-resource system, we further introduce a top-down-like control mechanism that serves to block overexploitation of the common resource. This control mechanism is not only motivated by theoretical works or lab experiments \cite{sugiarto_jpc17,fehr_aer00,sasaki_pnas12}, but is also stimulated by realistic field investigations focusing on the forest commons management \cite{yang_pnas13,rustagi_s10}. In contrast to a bottom-up self-regulation this top-down-type control assumes an effective external monitoring of agents. We have then shown that overexploitation is not the only danger of a sustainable state. In particular, it is not enough to restrict our share to a limit that is estimated from the actual abundance of the common resource, but we should also consider simultaneously the growing capacity of the latter. For example, if the growth rate is too small then even strong inspection and punishment are unable to prevent us from a depleted resource. Taking the other extreme, if the growth rate is high enough then the mentioned institutions have role only in how high resource level is stabilized, but we can always keep a sustainable state. As we pointed out monitoring overexploitation and punishing it has critical role at an intermediate growth rate of environmental resource, when efficient institutions can reverse the final destination of evolutionary process.

Renewable resources are generally believed to play key roles in achieving a sustainable human development \cite{valente_ere05}. Indeed, they are essential but we must consider not only temporal changes in environmental conditions but also their intrinsic features when designing their sustainable usage \cite{sugiarto_jpc17}. Otherwise our additional efforts to control participants become useless. Conceptually similar conclusion can also be obtained when ostracism is introduced as a collective control mechanisms into another coupled social-resource model proposed by Tavoni et al. \cite{tavoni_jtb12}. It is found that the stationary state of cooperators and defectors does not only depend on the ostracism strength, but also on the resource inflow \cite{sugiarto_jpc17}. We thus conclude that when we design a social control mechanism to solve the overexploitation problem in a coupled social-resource system we should first pay attention to the intrinsic features of the system which can determine in advance whether the designed control mechanism is capable to drive the level of common resource toward the desired direction.

Finally, we note that our model of dynamical cooperation and renewable environmental resource uses several simplifying assumptions and is just a first step toward a more sophisticated coevolutionary model. Still, we strongly believe that our results could be helpful to understand the role of growing capacity of renewable resources in designing control mechanism of punishment for governing the commons.

On the other hand, we may consider to relax or extend these simplifying assumptions for future research in order to understand better the coevolutionary dynamics of renewable resources and human cooperation in specific conditions. More precisely, in the present work we have considered that individual payoff in the replicator equation is identified as individual income originated directly from the common-pool resource. Indeed, there should exist a transformation relationship between individual income and payoff, which can be reflected by the production function \cite{tavoni_jtb12}. Thus incorporating such production function into our current model could be a further step toward a more realistic description. Meanwhile it is also interesting to study whether such transformation relationship can influence the evolutionary outcome in the coevolutionary framework we proposed. Second, as pointed out in Ref. \cite{weitz_pnas16} there exists a relative speed by which human behaviors modify the resource state, which was not considered in the framework of present model. Hence, a possible extension of our work could be to consider the relative changing speed between the resource dynamics and the cooperation level in the population. Third, we have considered a well-mixed interaction between agents where the environmental feedback was valid globally. Evidently, one may consider a structured population where both interactions and feedback from environment act locally, hence opening a new research path toward more realistic situations.

\section*{Supporting Information}
SI Text. Supporting Information for: Punishment and inspection for governing the commons in a feedback-evolving game.

\nolinenumbers

%
%

\begin{thebibliography}{10}

\bibitem{ostrom_90}
Ostrom E.
\newblock Governing the commons: The Evolution of Institutions for Collective
  Action.
\newblock Cambridge, U.K.: Cambridge University Press; 1990.

\bibitem{hardin_g_s68}
Hardin G.
\newblock The Tragedy of the Commons.
\newblock Science. 1968; 162:1243--1248.

\bibitem{pauly_n02}
Pauly D, Christensen V, Gu\'enette S, Pitcher TJ, Sumaila UR, Walters CJ,
  et~al.
\newblock Towards sustainability in world fisheries.
\newblock Nature. 2002; 418:689--695.

\bibitem{kraak_ff11}
Kraak SBM.
\newblock Exploring the 'public goods game' model to overcome the tragedy of
  the commons in fisheries management.
\newblock Fish and Fisheries. 2011; 12:18--33.

\bibitem{crespi_tee01}
Crespi BJ.
\newblock The evolution of social behavior in microorganisms.
\newblock Trends Ecol Evol. 2001; 16:178--183.

\bibitem{hummert_jtb10}
Hummert S, Hummert C, Schr{\"o}ter A, Hube B, Schuster S.
\newblock Game theoretical modelling of survival strategies of
  \protect{Candida} albicans inside macrophages.
\newblock J Theor Biol. 2010; 264:312--318.

\bibitem{west_prslb03}
West SA, Buckling A.
\newblock Cooperation, virulence and siderophore production in bacterial
  parasites.
\newblock Proc R Soc Lond B. 2003; 270:37--44.

\bibitem{rankin_tee07}
Rankin DJ, Bargum K, Kokko H.
\newblock The tragedy of the commons in evolutionary biology.
\newblock Trends Ecol Evol. 2007; 22:643--651.

\bibitem{gore_n09}
Gore J, Youk H, van Oudenaarden A.
\newblock Snowdrift game dynamics and facultative cheating in yeast.
\newblock Nature. 2009; 459:253--256.

\bibitem{schuster_bs11}
Schuster S, de~Figueiredo LF, Schroeter A, Kaleta C.
\newblock Combining Metabolic Pathway Analysis with Evolutionary Game Theory.
  Explaining the occurrence of low-yield pathways by an analytic optimization
  approach.
\newblock BioSystems. 2011; 105:147--153.

\bibitem{cox_es10}
Cox M, Arnold G, Tom{\'a}s SV.
\newblock A Review of Design Principles for Community-based Natural Resource
  Management.
\newblock Ecol Soc. 2012; 15:38.

\bibitem{hauser_n14}
Hauser OP, Rand DG, Peysakhovich A, Nowak MA.
\newblock Cooperating with the future.
\newblock Nature. 2014; 511:220-223.

\bibitem{sugiarto_prl17}
Sugiarto HS, Lansing JS, Chung NN, Lai CH, Cheong SA, Chew LY.
\newblock Social Cooperation and Disharmony in Communities Mediated through
  Common Pool Resource Exploitation.
\newblock Phys Rev Lett. 2017; 118:208301.

\bibitem{brander_aer98}
Brander J, Taylor MS.
\newblock The simple economics of easter island: A Ricardo-Malthus model of renewable resource use.
\newblock Am Econ Rev. 1998; 88:119--138.

\bibitem{tavoni_jtb12}
Tavoni A, Schl{\"u}ter M, Levin S.
\newblock The survival of the conformist: Social pressure and renewable
  resource management.
\newblock J Theor Biol. 2012; 299:152--161.

\bibitem{weitz_pnas16}
Weitz JS, Eksin C, Paarporn K, Brown SP, Ratcliff WC.
\newblock An oscillating tragedy of the commons in replicator dynamics with
  game-environment feedback.
\newblock Proc Natl Acad Sci USA. 2017; 113:E7518--E7525.

\bibitem{lade_te13}
Lade SJ, Tavoni A, Levin SA.
\newblock Regime shifts in a social ecological system.
\newblock Theor. Ecol. 2013; 6:359-372.

\bibitem{schluter_prsb16}
Schl\"{u}ter M, Tavoni A, Levin S.
\newblock Robustness of norm-driven cooperation in the commons.
\newblock Proc R Soc Lond B. 2016; 283: 20152431.

\bibitem{lee_jtb17}
Lee JH, Jusup M, Iwasa Y.
\newblock Games of corruption in preventing the overuse of common-pool
  resources.
\newblock J Theor Biol. 2017; 428:76--86.

\bibitem{fehr_aer00}
Fehr E, {G\"a}chter S.
\newblock Cooperation and Punishment in Public Goods Experiments.
\newblock Am Econ Rev. 2000; 90:980--994.

\bibitem{henrich_s06}
Henrich J.
\newblock Cooperation, Punishment, and the Evolution of Human Institutions.
\newblock Science. 2006; 312:60--61.

\bibitem{gachter_s08}
G\"{a}chter S, Renner E, Sefton M.
\newblock The long-run benefits of punishment.
\newblock Science. 2008; 687:1510.

\bibitem{sigmund_n10}
Sigmund K, De~Silva H, Traulsen A, Hauert C.
\newblock Social learning promotes institutions for governing the commons.
\newblock Nature. 2010; 466:861--863.

\bibitem{vasconcdlos_ncc13}
Vasconcelos VV, Santos FC, Pacheco JM.
\newblock A bottom-up institutional approach to cooperative governance of risky
  commons.
\newblock Nat Clim Change. 2013; 3:797--801.

\bibitem{pacheco_plr14}
Pacheco JM, Vasconcelos VV, Santos FC.
\newblock Climate change goverance, cooperation and self-organization.
\newblock Phys Life Rev. 2014; 11:573--586.

\bibitem{johnson_rsos15}
Johnson S.
\newblock Escaping the tragedy of the commons through targeted punishment.
\newblock R Soc Open Sci. 2015; 2:150223.

\bibitem{chen_jrsi15}
Chen X, Sasaki T, Br{\"a}nnstr{\"o}m {\AA}, Dieckmann U.
\newblock First carrot, then stick: how the adaptive hybridization of
  incentives promotes cooperation.
\newblock J R Soc Interface. 2015; 12:20140935.

\bibitem{chen_srep15}
Chen X, Sasaki T, Perc M.
\newblock Evolution of public cooperation in a monitored society with
  implicated punishment and within-group enforcement.
\newblock Sci Rep. 2015; 5:17050.

\bibitem{okada_ploscb15}
Okada I, Yamamoto H, Toriumi F, Sasaki T.
\newblock The effect of incentives and meta-incentives on the evolution of cooperation.
\newblock PLOS Comp Biol. 2015; 11:e1004232.

\bibitem{perc_pr17}
Perc M, Jordan JJ, Rand DG, Wang Z, Boccaletti S, Szolnoki A.
\newblock Statistical physics of human cooperation.
\newblock Phys Rep. 2017; 687:1--51.

\bibitem{wang_amc18}
Wang Q, He N, Chen X.
\newblock Replicator dynamics for public goods game with resource allocation in large populations.
\newblock Appl. Math. Comput. 2018; 328: 162-170.

\bibitem{nakamaru_pone14}
Nakamaru M, Yokoyama A.
\newblock The Effect of Ostracism and Optional Participation on the Evolution
  of Cooperation in the Voluntary Public Goods Game.
\newblock PLoS ONE. 2014; 9:e108423.

\bibitem{sugiarto_jpc17}
Sugiarto HS, Chung NN, Lai CH, Cheong SA, Chew LY.
\newblock Emergence of cooperation in a coupled socio-ecological system through
  a direct or an indirect social control mechanism.
\newblock J Phys Commun. 2017; 1:055019.

\bibitem{yang_pnas13}
Yang W, Liu W, {n}a AV, Tuanmu MN, He G, Dietz T, et~al.
\newblock Nonlinear effects of group size on collective action and resource
  outcomes.
\newblock Proc Natl Acad Sci USA. 2013; 110:10916--10921.

\bibitem{hofbauer_98}
Hofbauer J, Sigmund K.
\newblock Evolutionary Games and Population Dynamics.
\newblock Cambridge, U.K.: Cambridge University Press; 1998.

\bibitem{tsoularis_mb02}
Tsoularis A, Wallace J.
\newblock Analysis of logistic growth models.
\newblock Math Biosci. 2002; 179:21--55.

\bibitem{chen_srep14}
Chen X, Perc M.
\newblock Excessive abundance of common resources deters social responsibility.
\newblock Sci Rep. 2014; 4:4161.

\bibitem{brandt_pone12}
Brandt G, Merico A, Vollan B, Schl{\"u}ter A.
\newblock Human Adaptive Behavior in Common Pool Resource Systems.
\newblock PLoS ONE. 2012; 7:e52763.


\bibitem{khalil_12}
Khalil HK.
\newblock Nonlinear Systems.
\newblock Prentice Hall, NJ; 1996.

\bibitem{maciejewski_ploscb14}
Maciejewshi W, Fu F, Hauert C.
\newblock Evolutionary game dynamics in populations with heterogeneous structures.
\newblock PLOS Comp Biol. 2014; 10: e1003567.

\bibitem{adami_plr16}
Adami C, Schossau J, Hintze A.
\newblock Evolutionary game theory using agent-based methods.
\newblock Phys Life Rev. 2016; 19:1-26.

\bibitem{wu_ploscb17}
Wu T, Wang L, Fu F.
\newblock Coevolutionary dynamics of phenotypic diversity and contingent cooperation.
\newblock PLOS Comp Biol. 2017; 13: e1005363.

\bibitem{santos_n08}
Santos FC, Santos MD, Pacheco JM.
\newblock Social diversity promotes the emergence of cooperation in public goods games.
\newblock Nature. 2008; 454:213-216.

\bibitem{tavoni_ncc13}
Tavoni A.
\newblock Game theory: Building up cooperation.
\newblock Nat Clim Change. 2013; 3:782--783.

\bibitem{szolnoki_plr14}
Szolnoki A.
\newblock The power of games: Comment on ``Climate change governance,
  cooperation and self-organization" by Pacheco, Vasconcelos and Santos.
\newblock Phys Life Rev. 2014; 11:589--590.

\bibitem{sasaki_pnas12}
Sasaki T, Br{\"a}nnstr{\"o}m A, Dieckmann U, Sigmund K.
\newblock The take-it-or-leave-it option allows small penalties to overcome
  social dilemmas.
\newblock Proc Natl Acad Sci USA. 2012; 109:1165--1169.

\bibitem{rustagi_s10}
Rustagi D, Engel S, Kosfeld M.
\newblock Conditional cooperation and costly monitoring explain sucess in
  forest commons management.
\newblock Science. 2010; 330:961--965.

\bibitem{valente_ere05}
Valente S.
\newblock Sustainable Development, Renewable Resources and Technological
  Progress.
\newblock Env Res Econ. 2005; 30:115--125.

\end{thebibliography}

\begin{thebibliography}{1}

\bibitem{khalil_12}
Khalil HK.
\newblock Nonlinear Systems.
\newblock Prentice Hall, NJ 1996.

\end{thebibliography}

\begin{flushleft}

\newpage

{\Large
\textbf\newline{SI text (Supporting Information): Punishment and inspection for governing the commons in a feedback-evolving game} 
}
\newline
\\
Xiaojie Chen\textsuperscript{1*} and
Attila Szolnoki\textsuperscript{2$\dagger$}
\\
\bigskip
\textbf{1} School of Mathematical Sciences, University of Electronic
Science and Technology of China, Chengdu 611731, China
\\
\textbf{2} Institute of Technical Physics and Materials Science, Centre for Energy Research, Hungarian Academy of Sciences, P.O. Box 49, H-1525 Budapest, Hungary
\\
\bigskip

* xiaojiechen@uestc.edu.cn\\
$\dagger$ szolnoki@mfa.kfki.hu

\end{flushleft}



In this text, we provide the theoretical analysis of the equilibrium points in the following special cases.

\section{For $0<r=e_c<e_d$}

In this case, we have $r=\frac{b_mN}{R_m}$. Accordingly, the equation system becomes
$$ \left\{
\begin{aligned}
\dot{x} & = x(1-x)(p\beta-\frac{y}{R_m}b_m\alpha) \\
\dot{y} & =-\alpha ry+\alpha rxy-\frac{ry^2}{R_m}.
\end{aligned}
\right.
$$
Then the system has two fixed points in the parameter space, which are $[0,0]$ and $[1,0]$, respectively. Here the largest eigenvalue of $J(0,0)$ is positive, thus the fixed point $[0,0]$ is unstable.

For the fixed point $[1,0]$, we have
\begin{eqnarray*}
\begin{array}{cc}
J(1,0)=\left[\begin{array}{cc} -p\beta & 0 \\
0 & 0\end{array}\right].
\end{array}
\end{eqnarray*}
Then we cannot determine the stability of this fixed point based only on the eigenvalues of the Jacobian. Instead, we study its stability by further using the center manifold theorem \cite{khalil_12}. To do that, we take $x-1=z$, then the equation system becomes
$$ \left\{
\begin{aligned}
\dot{y} & = \alpha ryz-\frac{ry^2}{R_m} \\
\dot{z} & =-p\beta z+\frac{\alpha b_m yz}{R_m}-p\beta z^2+\frac{\alpha b_m yz^2}{R_m}.
\end{aligned}
\right.
$$
Using the center manifold theorem, we have that $z=h(y)$ is a center manifold for the above system. We start to try $h(y)=O(|y|^2)$, which yields the reduced system
$$\dot{y}  = -\frac{ry^2}{R_m}+O(|y|^3).$$
Since the fraction $\frac{r}{R_m}$ is nonzero, the fixed point $y=0$ of the reduced system is unstable. Consequently, the fixed point $[1, 0]$ is unstable as well \cite{khalil_12}.

\section{For $0<e_c<r<e_d$ and $\alpha b_m(1-\frac{e_c}{r})=p \beta$}

In this case, we have $1 + \frac{1}{\alpha} - \frac{R_m r}{\alpha b_m N}+\frac{p \beta R_m r}{{\alpha}^2 b^2_m N}< 1$ and $R_m-\frac{Nb_m}{r}=\frac{p\beta R_m}{\alpha b_m}$. As a result, there are three fixed points in the parameter space, which are $[0, 0]$, $[1,0]$, and $[1, R_m-\frac{Nb_m}{r}]$, respectively.

The fixed points $[0, 0]$ and $[1, 0]$ are both unstable since the largest eigenvalues of $J(0,0)$ and $J(1,0)$ are both positive.

For the fixed point $[1, R_m-\frac{Nb_m}{r}]$, we have
\begin{eqnarray*}
\begin{array}{cc}
J(1, R_m-\frac{Nb_m}{r})=\left[\begin{array}{cc} 0 & 0 \\
Np\beta & \frac{Nb_m-R_mr}{R_m}\end{array}\right],
\end{array}
\end{eqnarray*}
in which one eigenvalue is zero and the other eigenvalue is negative. Accordingly, we study its stability by further using the center manifold theorem \cite{khalil_12}. To do that, we take $x-1=z$ and $w=y-(R_m-\frac{N b_m}{r})$, then the equation system becomes
$$ \left\{
\begin{aligned}
\dot{z} & = -z(z+1)[p\beta-\frac{\alpha b_m(rw+R_mr-Nb_m)}{R_mr}] \\
\dot{w} & =\frac{[r(R_m+w)-Nb_m](\alpha b_m Nz-rw)}{R_m r}.
\end{aligned}
\right.
$$
We further let $M$ be a matrix whose columns are the eigenvectors of $J(1, R_m-\frac{Nb_m}{r})$, which can be written as
\begin{eqnarray*}
\begin{array}{cc}
M=\left[\begin{array}{cc} \frac{R_m r-Nb_m}{Np\beta R_m} & 0 \\
1& 1\end{array}\right].
\end{array}
\end{eqnarray*}

Then we have
\begin{eqnarray*}
\begin{array}{cc}
M^{-1}J(1, R_m-\frac{Nb_m}{r})M=\left[\begin{array}{cc} 0 & 0 \\
0& \frac{Nb_m-R_m r}{R_m r}\end{array}\right].
\end{array}
\end{eqnarray*}

We further take $[u \quad v]^T=M^{-1}[z \quad w]^T$, and thus we have $u=\frac{Np\beta R_m}{R_mr-Nb_m}z$ and $v=w-u$. It leads to
$$\dot{u} =\frac{\alpha b_m}{R_m}uv+\frac{\alpha b_m}{R_m}u^2+\frac{ru^2v}{R_mN}+\frac{ru^3}{R_mN}.$$

Using the center manifold theorem, we have that $v=h(u)$ is a center manifold. We start to try $h(u)=O(|u|^2)$, which yields the reduced system
$$\dot{u}  = \frac{\alpha b_m}{R_m}u^2+\frac{u^3}{R_mN}+O(|u|^4).$$

Since $\frac{\alpha b_m}{R_m}\neq 0$, the fixed point $u=0$ of the reduced system is unstable. Consequently, the fixed point $[1, R_m-\frac{Nb_m}{r}]$ of the original system is unstable \cite{khalil_12}.

\section{For $0<e_c<r=e_d$}

In this case, we have $1+\frac{1}{\alpha}-\frac{R_mr}{\alpha b_m N}=0$. In dependence of the efficiency of inspection and punishment we can further distinguish three following subcases.

($1$) For $b_m \alpha(1-\frac{e_c}{r})<p\beta$

We then have $1+\frac{1}{\alpha}-\frac{R_mr}{\alpha b_m N}+\frac{p\beta
R_mr}{{\alpha}^2 b^2_m N}=\frac{p\beta
R_mr}{{\alpha}^2 b^2_m N}>1$. As a result, the system has three fixed points in the parameter space. They are $[0, 0]$, $[1, 0]$, and $[1, R_m-\frac{Nb_m}{r}]$, respectively. According to the sign of the largest eigenvalues in the Jacobian matrices,  $[0,0]$ and $[1, 0]$ are unstable, while $[1,R_m-\frac{Nb_m}{r}]$ is a stable fixed point.

($2$) For $b_m \alpha(1-\frac{e_c}{r})>p\beta$

We then have $1+\frac{1}{\alpha}-\frac{R_mr}{\alpha b_m N}+\frac{p\beta
R_mr}{{\alpha}^2 b^2_m N}=\frac{p\beta
R_mr}{{\alpha}^2 b^2_m N}<1$. As a result, the system has four fixed points in the parameter space. They are $[0, 0]$, $[1, 0]$, $[1, R_m-\frac{Nb_m}{r}]$, and $[\frac{p\beta
R_mr}{{\alpha}^2 b^2_m N}, \frac{p\beta R_m}{\alpha b_m}]$, respectively. The first three fixed points are all unstable since the sign of the largest eigenvalues of the Jacobian matrices are all positive, while the last fixed point is stable because the term $\frac{p\beta R_m r}{\alpha b^2_mN}-\frac{R_mr}{Nb_m}+1$ is negative.

($3$) For $b_m \alpha(1-\frac{e_c}{r})=p\beta$

We then have $1+\frac{1}{\alpha}-\frac{R_mr}{\alpha b_m N}+\frac{p\beta
R_mr}{{\alpha}^2 b^2_m N}=\frac{p\beta
R_mr}{{\alpha}^2 b^2_m N}=1$ and $R_m-\frac{Nb_m}{r}=\frac{p\beta R_m}{\alpha b_m}$. As a result, the system has three fixed points in the parameter space. They are $[0, 0]$, $[1, 0]$, and $[1, \frac{p\beta R_m}{\alpha b_m}]$, respectively.

The fixed points $[0, 0]$ and $[1, 0]$ are both unstable since the largest eigenvalues of $J(0,0)$ and $J(1,0)$ are both positive.

For the fixed point $[1, \frac{p\beta R_m}{\alpha b_m}]$, we have
\begin{eqnarray*}
\begin{array}{cc}
J(1, \frac{p\beta R_m}{\alpha b_m})=\left[\begin{array}{cc} 0 & 0 \\
Np\beta & -\frac{p\beta r}{\alpha b_m}\end{array}\right],
\end{array}
\end{eqnarray*}
in which one eigenvalue is zero and the other eigenvalue is negative. Accordingly, we study its stability by further using the center manifold theorem \cite{khalil_12}. To do that, we take $x-1=z$ and $w=y-\frac{p\beta R_m}{\alpha b_m}$, then the equation system becomes
$$ \left\{
\begin{aligned}
\dot{z} & = \frac{\alpha}{R_m} b_mz(z+1)w \\
\dot{w} & =-\frac{r}{R_m}(w+\frac{p\beta R_m}{\alpha b_m})^2+\frac{N\alpha b_m}{R_m}(z+1)(w+\frac{p\beta R_m}{\alpha b_m}).
\end{aligned}
\right.
$$
We further let $M$ be a matrix whose columns are the eigenvectors of $J(1, \frac{p\beta R_m}{\alpha b_m})$, which can be written as
\begin{eqnarray*}
\begin{array}{cc}
M=\left[\begin{array}{cc} \frac{r}{\alpha b_m N} & 0 \\
1& 1\end{array}\right].
\end{array}
\end{eqnarray*}

Then we have
\begin{eqnarray*}
\begin{array}{cc}
M^{-1}J(1, \frac{p \beta r}{\alpha b_m})M=\left[\begin{array}{cc} 0 & 0 \\
0 & -\frac{p \beta r}{\alpha b_m} \end{array}\right].
\end{array}
\end{eqnarray*}

We further take $[u \quad v]^T=M^{-1}[z \quad w]^T$, and thus we have $u=\frac{\alpha b_m N}{r}z$ and $v=w-u$. It leads to
$$\dot{u} =\frac{\alpha b_m}{R_m}uv+\frac{\alpha b_m}{R_m}u^2+\frac{ru^2v}{R_mN}+\frac{ru^3}{R_mN}.$$

Using the center manifold theorem, we have that $v=h(u)$ is a center manifold. We start to try $h(u)=O(|u|^2)$, which yields the reduced system
$$\dot{u}  = \frac{\alpha b_m}{R_m}u^2+\frac{u^3}{R_mN}+O(|u|^4).$$

Since $\frac{\alpha b_m}{R_m}\neq 0$, the fixed point $u=0$ of the reduced system is unstable. Consequently, the fixed point $[1, \frac{p \beta r}{\alpha b_m}]$ of the original system is unstable \cite{khalil_12}.

\section{For $0<e_c<e_d<r$ and $\alpha b_m(1-\frac{e_c}{r})=p \beta$}

In this case, we have $1+\frac{1}{\alpha}-\frac{R_mr}{\alpha b_m N}+\frac{p\beta
R_mr}{{\alpha}^2 b^2_m N}=1$ and $R_m-\frac{Nb_m}{r}=\frac{p\beta R_m}{\alpha b_m}$. As a result, there are four fixed points, which are $[0, 0]$, $[1, 0]$, $[0, R_m-\frac{Nb_m(1+\alpha)}{r}]$, and $[1, R_m-\frac{Nb_m}{r}]$, respectively. The first three fixed points are all unstable since the sign of the largest eigenvalues of the Jacobian matrices are all positive. In addition, based on the theoretical analysis in Sec. $2$ of this text, we can conclude that the fixed point $[1, R_m-\frac{Nb_m}{r}]$ is also unstable.

\section{For $0<e_c<e_d<r$ and $\alpha b_m(1-\frac{e_d}{r})=p \beta$}

In this case, we have $1+\frac{1}{\alpha}-\frac{R_mr}{\alpha b_m N}+\frac{p\beta
R_mr}{{\alpha}^2 b^2_m N}=0$ and $R_m-\frac{N b_m (1+\alpha)}{r}=\frac{p\beta R_m}{\alpha b_m}$. As a result, the system has four fixed points in the parameter space. They are $[0, 0]$, $[1, 0]$, $[1, R_m-\frac{Nb_m}{r}]$, and $[0, \frac{p\beta R_m}{\alpha b_m}]$, respectively.

The fixed points $[0, 0]$, $[1, 0]$, and $[1, R_m-\frac{Nb_m}{r}]$ are all unstable since the largest eigenvalues of $J(0,0)$, $J(1,0)$, and $J(1, R_m-\frac{Nb_m}{r})$ are all positive.

For the fixed point $[0, \frac{p\beta R_m}{\alpha b_m}]$, we have
\begin{eqnarray*}
\begin{array}{cc}
J(0, \frac{p\beta R_m}{\alpha b_m})=\left[\begin{array}{cc} 0 & 0 \\
Np\beta & -\frac{p\beta r}{\alpha b_m}\end{array}\right],
\end{array}
\end{eqnarray*}
in which one eigenvalue is zero and the other eigenvalue is negative. Accordingly, we study its stability by using again the center manifold theorem \cite{khalil_12}. To do that, we take $x=z$ and $w=y-\frac{p\beta R_m}{\alpha b_m}$, then the equation system becomes
$$ \left\{
\begin{aligned}
\dot{z} & = \frac{\alpha b_m}{R_m} z(z-1)w \\
\dot{w} & =-Nb_m(1+\alpha-\alpha z)(\frac{w}{R_m}+\frac{p\beta}{\alpha b_m})+r(w+\frac{p\beta R_m}{\alpha b_m})(1-\frac{w}{R_m}-\frac{p\beta}{\alpha b_m}).
\end{aligned}
\right.
$$
We further let $M$ be a matrix whose columns are the eigenvectors of $J(0, \frac{p\beta R_m}{\alpha b_m})$, which can be written as
\begin{eqnarray*}
\begin{array}{cc}
M=\left[\begin{array}{cc} \frac{r}{\alpha b_m N} & 0 \\
1& 1\end{array}\right].
\end{array}
\end{eqnarray*}

Then we have
\begin{eqnarray*}
\begin{array}{cc}
M^{-1}J(0, \frac{p \beta r}{\alpha b_m})M=\left[\begin{array}{cc} 0 & 0 \\
0 & -\frac{p \beta r}{\alpha b_m} \end{array}\right].
\end{array}
\end{eqnarray*}

We further take $[u \quad v]^T=M^{-1}[z \quad w]^T$, and thus we have $u=\frac{\alpha b_m N}{r}z$ and $v=w-u$. It leads to
$$\dot{u} =-\frac{r}{R_m N}uv-\frac{r}{R_mN}u^2+\frac{r^2u^2v}{R_m\alpha b_mN^2}+\frac{r^2u^3}{R_m\alpha b_mN^2}.$$

Using the center manifold theorem, we have that $v=h(u)$ is a center manifold. We start to try $h(u)=O(|u|^2)$, which yields the reduced system
$$\dot{u}  = -\frac{r}{R_mN}u^2+\frac{r^2u^3}{R_m\alpha b_mN^2}+O(|u|^4).$$

Since the fraction $\frac{r}{R_mN}$ is nonzero, the fixed point $u=0$ of the reduced system is unstable. Consequently, the fixed point $[0, \frac{p\beta R_m}{\alpha b_m}]$ of the original system is unstable \cite{khalil_12}.

\nolinenumbers

%
%
%

\end{document}